\documentclass[prb,superscriptaddress,showpacs,floatfix,tightenlines,10pt,twocolumn]{revtex4-1}
\usepackage{amssymb}
\usepackage{amsmath}
\usepackage{graphicx}
\usepackage{float}
\usepackage{xcolor}

\begin{document}

\title{Quantum magnetic oscillations in Weyl semimetals with tilted nodes}
\author{Samuel Vadnais}
\author{Ren\'{e} C\^{o}t\'{e}}
\affiliation{D\'{e}partement de physique and Institut Quantique, Universit\'{e} de
Sherbrooke, Sherbrooke, Qu\'{e}bec, J1K 2R1, Canada }
\date{\today }

\begin{abstract}
A Weyl semimetal (WSM)\ is a three-dimensional topological phase of matter
where pairs of nondegenerate bands cross at isolated points in the Brillouin
zone called Weyl nodes. Near these points, the electronic dispersion is
gapless and linear. A magnetic field $B$ changes this dispersion into a set
of positive and negative energy Landau levels which are dispersive along the
direction of the magnetic field only. In this set, the $n=0$ Landau level is
special since its dispersion$\ $is linear and unidirectional. The presence
of this chiral level distinguishes Weyl from Schr\"{o}dinger fermions. In
this paper, we study the quantum oscillations of the orbital magnetization
and magnetic susceptibility in Weyl semimetals. We generalise earlier works%
\cite{Mikitik2019} on these De Haas-Van Alphen oscillations by considering
the effect of a tilt of the Weyl nodes. We study how the fundamental period
of the oscillations in the small $B$ limit and the strength of the magnetic
field $B_{1}$ required to reach the quantum limit (i.e. where the Fermi
level is lying in the chiral level) are modified by the magnitude and
orientation of the tilt vector $\mathbf{t}$. We show that the magnetization
from a single node is finite in the $B\rightarrow 0$ limit. Its sign depends
on the product of the chirality and sign of the tilt component along the
magnetic field direction. We also study the magnetic oscillations from a
pair of Weyl nodes with opposite chirality and with opposite or identical
tilt. Our calculation shows that these two cases lead to a very different
behavior of the magnetization in the small and large $B$ limits. We finally
consider the effect of an energy shift $\pm \Delta _{0}$ of a pair of Weyl
nodes on the magnetic oscillations. We assume a constant density of carriers
so that both nodes share a common Fermi level and the density of carriers is
constantly redistributed between the two nodes as the magnetic field is
varied. Our calculation can easily be extended to a WSM with an arbitrary
number of pairs of Weyl nodes.
\end{abstract}

\maketitle

\section{INTRODUCTION}

A Weyl semimetal\cite{WSM review} (WSM) is a three-dimensional topological
phase of matter where pairs of nondegenerate bands cross at isolated points
in the Brillouin zone called Weyl nodes. Near these points, the electronic
dispersion is gapless and linear in momentum and the excitations satisfy the
Weyl equation, a two-component analog of the Dirac equation. Each Weyl node
has a chirality index $\chi ,$ an integer reflecting the topological nature
of the band structure. For the Weyl points to be stable, either
time-reversal or inversion symmetry or both must be broken so that the two
bands that cross are nondegenerate.

Weyl semimetals show a number of interesting transport properties, such as
an anomalous Hall effect\cite{AHE} for a WSM\ with broken time-reversal
symmetry, a chiral-magnetic effect\cite{CME} for Weyl semimetals that break
inversion symmetry, gapless surface states called Fermi arcs\cite{Fermi arcs}
and a chiral anomaly leading to a negative longitudinal magnetoresistance%
\cite{Chiral anomaly}.

A magnetic field replaces the linear dispersion by a set of positive ($n>0$)
and negative ($n<0$) energy levels. These Landau levels are dispersive along
the direction of the magnetic field. For $n\neq 0$ and in the simplest case
(no tilt or energy shift of the nodes), the energy of each level is $%
E_{n\neq 0}\left( k\right) =\left( \hslash v_{F}/\ell \right) $sgn$(n)\sqrt{%
k^{2}\ell ^{2}+2\left\vert n\right\vert }$, where $k$ is a wave vector in
the direction of the magnetic field, $v_{F}$ is the Fermi velocity and $\ell
=\sqrt{\hslash /eB}$ is the magnetic length with $B$ the magnetic field. The 
$n=0$ Landau level is special since its dispersion$\ $is linear,
unidirectional and independent of the magnetic field i.e. $E_{n=0}\left(
k\right) =-\chi \hslash v_{F}k$, where $\chi $ is the chirality index. The
presence of this chiral level affects many properties of Weyl semimetals
such as the optical absorption spectrum which is different from that of Schr%
\"{o}dinger or Dirac fermions\cite{MagnetoSigma,Bertrand2019} or the Faraday
and Kerr effects\cite{Parent2020,Randeria,Levy}.

The magnetic susceptibility of Weyl semimetals also shows unusual
characteristics such as a diverging diamagnetic susceptibility when the
chemical potential is close to the neutrality point in the limit $%
B\rightarrow 0,$ a spontaneous magnetization in this limit if the nodes are
tilted in momentum space and a phase shift of the De Haas-Van alphen
oscillations with respect to those due to Schr\"{o}dinger fermions. The
magnetic susceptibility of Weyl and Dirac semimetals (and more generally
near points in the Brillouin zone of crystals where bands are degenerate\cite%
{Mikitik1989,Mikitik2021}) has been studied by a number of authors. A recent
review (up to the year 2019) is given in Ref. \onlinecite{Mikitik2019}.

In the present paper, we complement these earlier works by considering Weyl
nodes which are shifted in energy and/or tilted in momentum space. We study
the contribution of the added electrons or holes to the orbital
magnetization and magnetic susceptibility. It has been shown before that a
tilt modifies the dynamical conductivity\cite{Carbotte3} and the selection
rules for electromagnetic absorption\cite{Goerbig2016}. It can lead to
interesting effects such as providing a signature of the valley polarization%
\cite{Bertrand2019} and the chiral anomaly\cite{Parent2020}, induces
dichroism\cite{Carbotte2} and an anisotropic chiral magnetic effect\cite%
{Wurff}. In the present work, we show that a tilt modifies the behavior of
the quantum oscillations of the orbital magnetization and magnetic
susceptibility and renders them anisotropic with respect to the orientation
of the tilt vector. We use a mostly numerical approach so that we can
compute these oscillations for an arbitrary magnetic field. We discuss the
period $P$ of the oscillations in the small magnetic field limit (i.e. the
fundamental period) as well as the value of the magnetic field $B_{1}$
required to reach the quantum limit where the Fermi level is lying in the
chiral $n=0$ Landau level. Both quantities can be measured by torque
magnetometry experiments\cite{Moll,Modic}. For a single Weyl node, the
magnetization is finite in the $B\rightarrow 0$ limit and its sign depends
on the product of the chirality $\chi =\pm 1$ and sign of the component of
the tilt along the magnetic field direction $t_{z}.$ Hence, at least two
nodes with opposite values of the product $\chi t_{z}$ are necessary for the
magnetization to vanish in the classical ($B=0$) limit as required on
physical ground.

After studying the single node case, we consider the magnetic oscillations
from a pair of Weyl nodes with opposite chirality. We compute the magnetic
oscillations for two nodes with the same or opposite value of the tilt
component $t_{z}.$ Since the density of states is not the same for positive
or negative value of $t_{z},$ the density of carriers in each node is also
different for a given Fermi level. Indeed, the total density of carriers
(electrons minus holes, measured with respect to the vacuum state), and not
the chemical potential, is fixed in our calculation, so that the two nodes
share a common Fermi level. The density of carriers in each node is
constantly readjusted as the magnetic field is varied to produce the quantum
oscillations. This reequilibration of the carrier density and the dependence
of the fundamental period on the tilt vector leads to a complex behavior for
the magnetic oscillations. We complete our study by discussing the behavior
of the oscillations from a pair of Weyl nodes shifted in energy by a bias $%
\pm \Delta _{0}$ but untilted. For large $\Delta _{0},$ the density in the
two nodes can be made very different thus modifying more importantly the
pattern of the quantum oscillations.

Our paper is organized as follows. In Sec. II, we describe the formalism
needed to compute the magnetization and differential magnetic
susceptibility. We study the magnetic oscillations from a single node in
Sec. III and from a pair of Weyl nodes in Sec. IV. We conclude in Sec. V.

\section{FORMALISM}

\subsection{Landau levels for a WSM\ in a magnetic field}

The Hamiltonian for the electrons in a node of a WSM at wave vector $\mathbf{%
Q}_{\tau }$ in the Brillouin zone is given, for small wave vector $\mathbf{k}
$ measured from $\mathbf{Q}_{\tau }$ by%
\begin{equation}
h_{\tau }\left( \mathbf{k}\right) =\hslash v_{F,\tau }\left( -\chi _{\tau }%
\mathbf{\sigma }\cdot \mathbf{k}+Q_{0,\tau }\sigma _{0}+\mathbf{t}_{\tau
}\cdot \mathbf{k}\sigma _{0}\right) ,
\end{equation}%
where $\tau =1,2,3,...$ is the node index. Each node can have its own Fermi
velocity $v_{F,\tau },$ chirality $\chi _{\tau }$, energy bias $\Delta
_{0,\tau }=\hbar v_{F}Q_{0.\tau }$ and tilt $\mathbf{t}_{\tau }$ (a unitless
vector). In this equation, $\mathbf{\sigma }$ is a vector of Pauli matrices
in the $1/2$ pseudospin state of the bands at their crossing point and $%
\sigma _{0}$ is the $2\times 2$ unit matrix. We restrict our analysis to
type I WSMs where $\left\vert \mathbf{t}_{\tau }\right\vert <1$ and assume
that the energy bias $\Delta _{0,\tau }$ and the range of $\left\vert 
\mathbf{k}\right\vert $ are small enough for the dispersion to remain linear
so that we can work in the confine of the continuum model. Hereafter and
until Sec. IV, we study the quantum oscillations of a single node. We thus
drop the index $\tau $ to simplify the notation.

\begin{figure}
\centering
\includegraphics[width = \linewidth]{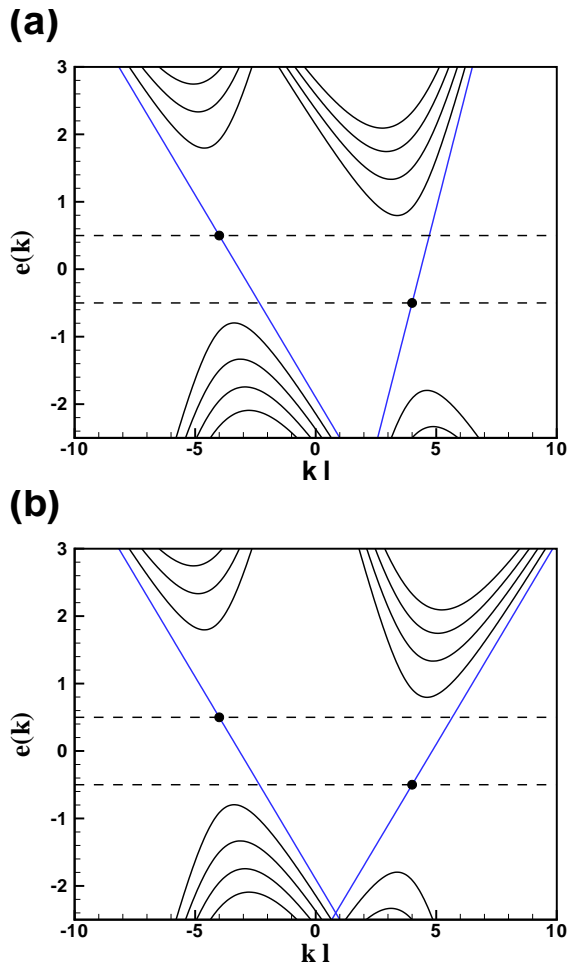} 
\caption{Energy in units of $\hslash
v_{F}/\ell $ for the first Landau levels for two nodes with opposite
chirality and bias. Parameters are $\protect\chi _{1}=-1,Q_{0,1}\ell =0.5$
for the node on the left and $\protect\chi _{2}=+1,Q_{0,2}\ell =-0.5$ for
the node on the right and : (a) $t_{1,z}=t_{2,z}=0.4$ and (b) $t_{1,z}=-t_{2,z}=0.4$. The blue lines and the black dots are the chiral
level and Dirac point in each node. The separation between the nodes is
arbitrary.} \label{fig1}
\end{figure}

In a magnetic field $\mathbf{B}=B_{0}\widehat{\mathbf{z}}$, the kinetic
energy is quantized into Landau levels with index $n=0,\pm 1,\pm 2,...$
Level $n=0$ is called the chiral Landau level and its dispersion is given by%
\cite{Mikitik1996,Yu2016,Udagawa2016,Goerbig2016}

\begin{equation}
e_{n=0}\left( k\right) =Q_{0}\ell +\left( t_{z}+\chi \beta \right) k\ell ,
\label{ezero}
\end{equation}%
where, from now on, $k$ is a wave vector along the magnetic field direction.
For Landau levels $n\neq 0$, the dispersion is%
\begin{eqnarray}
e_{n\neq 0}\left( k\right) &=&Q_{0}\ell +t_{z}k\ell  \label{esup} \\
&&+\text{sgn}(n)\beta \sqrt{k^{2}\ell ^{2}+2\beta \left\vert n\right\vert },
\notag
\end{eqnarray}%
where sgn is the signum function and we have defined 
\begin{eqnarray}
t_{z} &=&t\cos \theta , \\
t_{\bot } &=&t\sin \theta , \\
\beta &=&\sqrt{1-t_{\bot }^{2}},
\end{eqnarray}%
with $\ell =\sqrt{\hslash /eB_{0}}$ the magnetic length and $\theta \mathbf{%
\ }$the polar angle of the tilt vector. All energies are given in units of $%
\hslash v_{F}/\ell $ unless specified otherwise. The dispersion of the
Landau levels and the other physical quantities that we compute in this
paper do not depend on the azimuthal angle $\varphi $ of the tilt vector.
Figure \ref{fig1} shows the Landau level dispersion for a WSM\ with two
nodes of opposite chirality $\chi _{1}=-\chi _{2}=1$ and (unitless) bias $%
Q_{0,1}\ell =-Q_{0,2}\ell =0.5$ for :\ (a) same tilt $t_{1,z}=t_{2,z}=0.4$
and (b) opposite tilt $t_{1,z}=-t_{2,z}=0.4$ A finite value of $t_{\bot }$
(positive or negative) decreases the separation in energy between adjacent
Landau levels (not shown in the figure). A positive (negative) bias $%
Q_{0}\ell $ shifts the Landau levels upward (downward) in energy.

The minimal (maximal) energy in level $n>0$ ($n<0$) is given by%
\begin{eqnarray}
\min \left[ e_{n>0}\right] &=&Q_{0}\ell +\sqrt{2\beta \gamma n},
\label{emin} \\
\max \left[ e_{n<0}\right] &=&Q_{0}\ell -\sqrt{2\beta \gamma \left\vert
n\right\vert ,}  \label{emax2}
\end{eqnarray}%
where we have defined 
\begin{equation}
\gamma =1-t^{2}.
\end{equation}%
These extrema occur at wave vector%
\begin{equation}
\left( k\ell \right) _{\text{ext}}=-\text{sgn}(n)\sqrt{\frac{2\left\vert
n\right\vert \beta }{\gamma }}t_{z}.  \label{kmin}
\end{equation}

The energy bias in real energy units $\Delta _{0}$ is independent of the
magnetic field while the unitless energy bias $Q_{0}\ell $ varies with the
magnetic field according to the relation%
\begin{equation}
Q_{0}\ell =\frac{\Delta _{0}}{\hbar v_{F}/\ell }.  \label{q0}
\end{equation}%
The dispersion $E_{n=0}=\frac{\hslash v_{F}}{\ell }e_{n=0}$ of the chiral
level in real energy units is independent of the magnetic field.

\subsection{Density of states}

At energy $e,$ the level index of the highest partially occupied Landau
level in each node is%
\begin{equation}
n_{\max }\left( e\right) =\text{sgn}(e-Q_{0}\ell )\lfloor \frac{\left(
e-Q_{0}\ell \right) ^{2}}{2\beta \gamma }\rfloor ,
\end{equation}%
where $\lfloor \rfloor $ is the floor function.

\begin{figure}
\centering
\includegraphics[width = \linewidth]{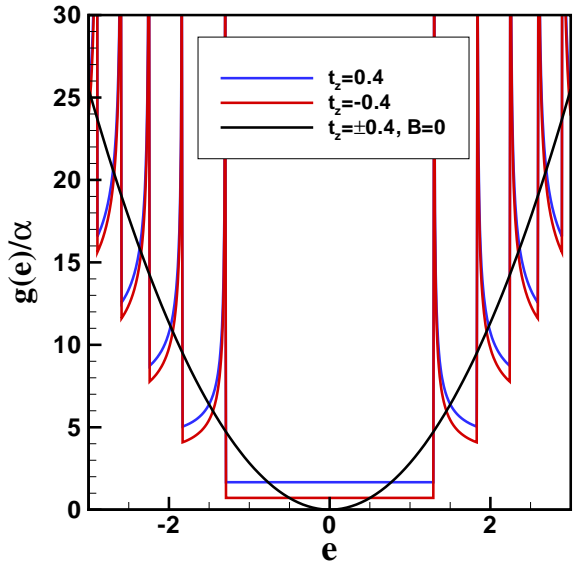} 
\caption{Density of states as a function
of the energy $e$ for a single node with zero bias, chirality $\protect\chi =-1$ and for $t_{z}=\pm 0.4.$ The black line is the $B=0,t_{z}=\pm 0.4$
result which does not depend on the sign of $t_{z}$. } \label{fig2}
\end{figure}

The density of states (DOS)\ $g\left( e\right) $ per unit volume $V$ is 
\begin{eqnarray}
g\left( e\right) &=&\frac{1}{V}N_{\varphi }\sum_{n,k}\delta \left( \frac{%
\hslash v_{F}}{\ell }\left( e-e_{n}\left( k\right) \right) \right)
\label{dos} \\
&=&\frac{\alpha }{\beta +\chi t_{z}}  \notag \\
&&+\sum_{n=1}^{n_{\max \left( e\right) }}\sum_{j=\pm 1}\frac{\alpha \Theta
\left( e-Q_{0}\ell \right) }{\left\vert t_{z}+\frac{\beta k_{n,j}\ell }{%
\sqrt{k_{n,j}^{2}\ell ^{2}+2\beta n}}\right\vert }  \notag \\
&&+\sum_{n=n_{\max }\left( e\right) }^{-1}\sum_{j=\pm 1}\frac{\alpha \Theta
\left( Q_{0}\ell -e\right) }{\left\vert t_{z}-\frac{\beta k_{n,j}\ell }{%
\sqrt{k_{n,j}^{2}\ell ^{2}+2\beta \left\vert n\right\vert }}\right\vert }, 
\notag
\end{eqnarray}%
\newline
where the constant $\alpha $ is defined by%
\begin{equation}
\alpha =\frac{1}{4\pi ^{2}\ell ^{3}}\frac{1}{\hslash v_{F}/\ell }.
\end{equation}%
(Note that $\beta +\chi t_{z}>0$ for all angles $\theta .$) Each Landau
level $\left( n,k\right) $ has a degeneracy given by $N_{\varphi }=S/2\pi
\ell ^{2},$ where $S$ is the area of the WSM\ perpendicular to the magnetic
field. In Eq. (\ref{dos}), the wave vectors $k_{n,\pm }\ell $ are defined by

\begin{eqnarray}
k_{n,\pm }\ell &=&-\frac{1}{\gamma }\left( e-Q_{0}\ell \right) t_{z}
\label{dos1} \\
&&\pm \frac{\beta }{\gamma }\sqrt{\left( e-Q_{0}\ell \right)
^{2}-2\left\vert n\right\vert \beta \gamma }.  \notag
\end{eqnarray}%
The $k_{n,\pm }\ell $ are the two $k$ points in each level $n\neq 0$ where $%
e_{n}\left( k_{n,\pm }\ell \right) =e_{F}$ with $e_{F}$ the unitless Fermi
level. At a band extremum, they merge into a single point with wave vector $%
k_{n,j=\pm }\ell =\left( k_{n}\ell \right) _{\text{ext}}$ given by Eq. (\ref%
{kmin}). At this particular point, the denominator in the third line of Eq. (%
\ref{dos}) goes to zero and the density of states diverges as shown in Fig. %
\ref{fig2}.

At zero tilt and bias, Eq. (\ref{dos}) reduces to the known result\cite%
{Carbotte1} : 
\begin{equation}
g\left( e\right) =\alpha \left[ 1+2\left\vert e\right\vert
\sum_{n=1}^{\left\vert n_{\max }\left( e\right) \right\vert }\frac{1}{\sqrt{%
e^{2}-2\left\vert n\right\vert }}\right] ,
\end{equation}%
and at zero magnetic field to:%
\begin{equation}
g\left( E\right) =\frac{1}{2\pi ^{2}}\frac{\left( E-\Delta _{0}\right) ^{2}}{%
\left( \hslash v_{F}\right) ^{3}}\frac{1}{\left( 1-t^{2}\right) ^{2}},
\label{dosclassique}
\end{equation}%
which is represented by the black line in Fig. \ref{fig2}.

The term in the second line of Eq. (\ref{dos}) is the contribution of the
chiral level to the density of states. It is independent of the energy but
increases linearly with the magnetic field. The density of states depends on
the chirality and tilt vector only through the product $\chi t_{z}$ only. As
for the contribution of the $n\neq 0$ levels, it can be deduced from Eq. (%
\ref{esup}) and the summation over $k$ in Eq. (\ref{dos}) that it is
independent of the sign of $t_{z}$ because of the symmetry relation $%
e_{n\neq 0}\left( k,t_{z},Q_{0}\ell \right) =e_{n\neq 0}\left(
-k,-t_{z},Q_{0}\ell \right) .$ It is also independent of the chirality
index. It is thus convenient to define the density of states for a node as
the sum of the two contributions:%
\begin{equation}
g\left( e\right) =g_{0,\chi }+g_{>}\left( e-Q_{0}\ell \right) ,
\label{dosnoeud}
\end{equation}%
where $g_{>}\left( e\right) $ is the density of states from levels $n\neq 0$
defined with $Q_{0}=0$ and%
\begin{equation}
g_{0,\chi }=\frac{\alpha }{\beta +\chi t_{z}}  \label{dosnoeud2}
\end{equation}%
is the contribution of the chiral level.

Figure \ref{fig2} shows the sawtooth behavior of the density of states as a
function of the unitless energy $e$ for a single node with zero bias,
chirality $\chi =-1$ and for $t_{z}=\pm 0.4.$ The density of states\ from
the chiral level is reduced (increased) from its $t_{z}=0$ value when $\chi
t_{z}>0$ ($\chi t_{z}<0).$ Equation (\ref{emin}) shows that the gap between
the positive and negative energy levels is reduced by a finite value of $%
\left\vert \mathbf{t}\right\vert $. A finite bias only shifts the function $%
g\left( e\right) $ globally to $e>0$ $\left( e<0\right) \,$ for $Q_{0}$
positive (negative). The separations between the square root singularities
in the density of states scale as $\sqrt{B}$ for a Weyl fermions in contrast
with three-dimensional Schr\"{o}dinger fermions where it increases linearly
with the magnetic field$.$

\subsection{Magnetization and magnetic susceptibility}

Throughout our paper, we work at $T=0$ K so that the magnetization \textit{%
per electron} in units of the Bohr magneton $\mu _{B}=e\hslash /2m_{e}$
(where $m_{e}$ is the bare electron mass) is obtained by taking the
derivative of the electronic energy per electron $U$ (which we define below)
with respect to the magnetic field \textit{at constant density}: 
\begin{equation}
m=-\frac{1}{\mu _{B}}\left. \frac{\partial U}{\partial B}\right\vert
_{n_{e}}.  \label{mag}
\end{equation}%
Differentiating the energy a second time gives the (differential) magnetic
susceptibility \textit{per electron }in units of Bohr magneton per Tesla:%
\begin{equation}
\chi _{m}=-\frac{1}{\mu _{B}}\left. \frac{\partial ^{2}U}{\partial B^{2}}%
\right\vert _{n_{e}}=\left. \frac{\partial m}{\partial B}\right\vert
_{n_{e}}.  \label{sus}
\end{equation}

\section{MAGNETIC SUSCEPTIBILITY FROM A SINGLE WEYL\ NODE}

In this section, we derive the magnetic oscillations from the electrons in a
single node. We can set $Q_{0}=0$ in all formulas since shifting the zero of
energy (the Dirac point)\ of a node when its density $n_{e}$ is fixed does
not change its magnetization or susceptibility.

\subsection{Fermi level and density of carriers}

The vacuum state is defined as the filled valence band of the Dirac cone. We
define the carrier density with respect to that vacuum state. It is positive
for electrons ($e_{F}>0$) and negative for holes $\left( e_{F}<0\right) .$
According to Eqs. (\ref{emin},\ref{emax2}), the Fermi level is in the chiral
level when $\left\vert e_{F}\right\vert <\sqrt{2\beta \gamma }$ and
intersects the Landau level $n\neq 0$ when%
\begin{equation}
\left\vert e_{F}\right\vert \geq \sqrt{2\left\vert n\right\vert \beta \gamma 
}.
\end{equation}%
The density of carriers is related to the chemical potential by the equation 
\begin{eqnarray}
n_{e} &=&\frac{\hslash v_{F}}{\ell }\int_{0}^{e_{F}}g\left( e\right) de
\label{densities} \\
&=&\frac{1}{4\pi ^{2}\ell ^{3}}\frac{e_{F}}{\chi t_{z}+\beta }  \notag \\
&&+\frac{\Theta \left( e_{F}\right) }{4\pi ^{2}\ell ^{3}}\sum_{n=1}^{n=n_{%
\max }\left( e_{F}\right) }\Lambda _{n}\left( e_{F}\right)  \notag \\
&&-\frac{\Theta \left( -e_{F}\right) }{4\pi ^{2}\ell ^{3}}\sum_{n=n_{\max
}\left( e_{F}\right) }^{n=-1}\Lambda _{n}\left( e_{F}\right) ,  \notag
\end{eqnarray}%
where we have defined%
\begin{eqnarray}
\Lambda _{n}\left( e\right) &=&k_{n,+}\ell \left( e\right) -k_{n,-}\ell
\left( e\right) \\
&=&2\frac{\beta }{\gamma }\sqrt{e^{2}-2\left\vert n\right\vert \beta \gamma }%
.  \notag
\end{eqnarray}%
The oscillations of the Fermi level $e_{F}\left( B\right) $ with magnetic
field are found by solving Eq. (\ref{densities}) with $n_{e}$ fixed$.$ A
numerical evaluation shows that, when $B\rightarrow 0,$ Eq. (\ref{densities}%
) reduces to the classical result 
\begin{equation}
E_{F}=\text{sgn}\left( n_{e}\right) \hslash v_{F}\left( 6\pi ^{2}\left(
1-t^{2}\right) ^{2}\left\vert n_{e}\right\vert \right) ^{1/3}.
\label{classical}
\end{equation}

\subsection{Electronic energy}

At zero temperature, the kinetic energy \textit{per carrier} is 
\begin{equation}
U=\frac{1}{\left\vert n_{e}\right\vert }\left( \frac{\hslash v_{F}}{\ell }%
\right) ^{2}\int_{0}^{e_{F}}g\left( e\right) ede.
\end{equation}%
It is positive for both electron ($n_{e}>0,e_{F}>0$) or hole ($%
n_{e}<0,e_{F}<0$) carriers. Using the definition of the density of states,
the energy becomes 
\begin{eqnarray}
U &=&\frac{1}{2}\zeta \frac{e_{F}^{2}}{\beta +\chi t_{z}}  \label{energy} \\
&&+\zeta \Theta \left( e_{F}\right) \sum_{n=1}^{n_{\max }\left( e_{F}\right)
}\int_{k_{n,-}\ell \left( e_{F}\right) }^{k_{n,+}\ell \left( e_{F}\right)
}e_{n}\left( x\right) dx  \notag \\
&&-\zeta \Theta \left( -e_{F}\right) \sum_{n=n_{\max }\left( e_{F}\right)
}^{n=-1}\int_{k_{n,-}\ell \left( e_{F}\right) }^{k_{n,+}\ell \left(
e_{F}\right) }e_{n}\left( x\right) dx,  \notag
\end{eqnarray}%
where we have defined 
\begin{equation}
\zeta =\frac{\hslash v_{F}/\ell }{4\pi ^{2}\left\vert n_{e}\right\vert \ell
^{3}}=0.385\frac{B^{2}}{\left\vert \overline{n}_{e}\right\vert }\overline{v}%
_{F}\text{ (meV)}
\end{equation}%
We define $\overline{n}_{e}$ and $\overline{v}_{F}$ as the unitless carrier
density and Fermi velocity by $n_{e}=\overline{n}_{e}\times 10^{22}$ m$^{-3}$
and $v_{F}=\overline{v}_{F}\times 10^{5}$ m/s. In our numerical calculation,
we use $\overline{v}_{F}=3$ and $\overline{n}_{e}=2$. For comparison, in the
Weyl semimetal TaAs, $\overline{v}_{F}\approx 3.6$ and $\overline{n}%
_{e}\approx 0.42$ for the W1 nodes and $\overline{n}_{e}\approx 0.00105$ for
the W2 nodes.

The integrals in Eq. (\ref{energy}) can be evaluated analytically to give%
\begin{eqnarray}
\int e_{n}\left( x\right) dx &=&\frac{1}{2}x^{2}t_{z}  \label{ep} \\
&&+\frac{1}{2}\text{sgn}(n)\beta x\sqrt{x^{2}+2\beta \left\vert n\right\vert 
}  \notag \\
&&+\beta ^{2}n\ln \left( x+\sqrt{x^{2}+2\beta \left\vert n\right\vert }%
\right) .  \notag
\end{eqnarray}%
Equation (\ref{energy}) reduces to the energy result given by Eq. (33) of
Ref. (\onlinecite{Carbotte1}) calculated in the absence of tilt and bias. At
equal density, the energy $U$ is the same for electron and hole carriers.
The magnetization and susceptibility are then also the same and we can,
without loss of generality, consider only electron carriers for the rest of
this section.

\begin{figure}
\centering
\includegraphics[width = \linewidth]{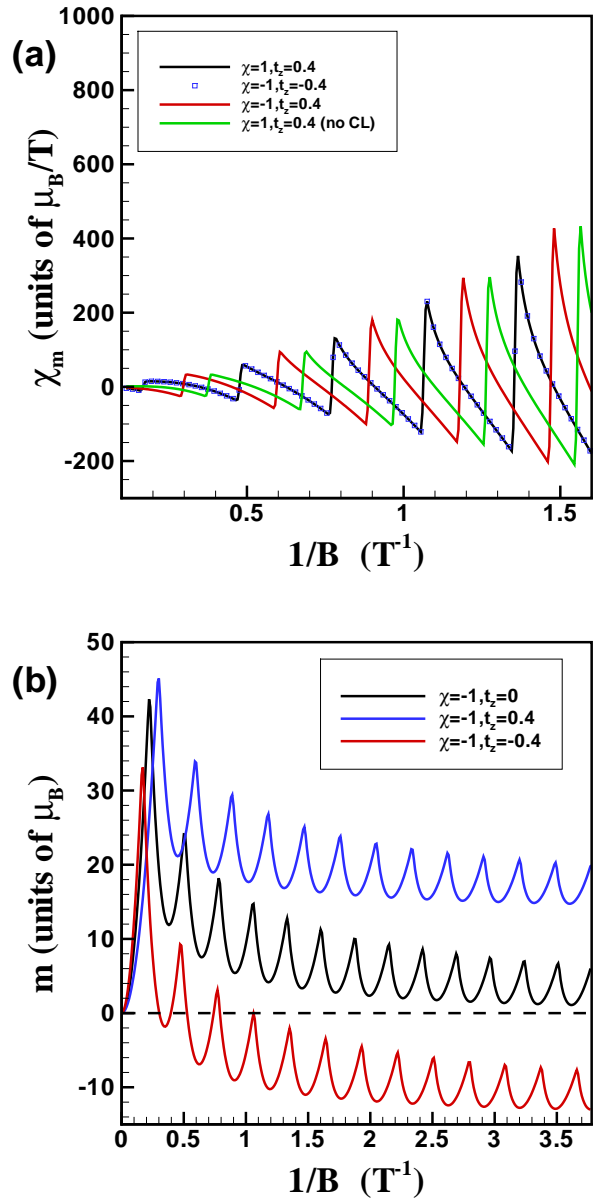} 
\caption{(a) Magnetic susceptibility at
zero bias from a single node for different chiralities and sign of the $z$
component of the tilt vector. The green line is for a node where the chiral
level has been artifically removed. The large $B$ behavior is shown in the
inset. (b) Magnetisation for a node with chirality -1 and tilt vector $t_{z}=0,0.4,-0.4$ showing the different behaviors in the small $B$ limit.}
\label{fig3}
\end{figure}

Figure \ref{fig3} shows an example of quantum oscillations of the magnetic
susceptibility and magnetization for $\chi =\pm 1,t_{z}=0,\pm 0.4$ and $%
\overline{n}_{e}=2.$ The oscillations are identical for two nodes with the
same sign of the product $\chi t_{z}$. For the susceptibility
(magnetization), they increase (decrease)\ in amplitude as $1/B$ increases.
Each discontinuity in the slope of the oscillations indicates a transition
of the Fermi level from $n$ to $n+1$ if $1/B$ increases. At high magnetic
field, the WSM\ enters the quantum regime where the Fermi level intersects
only the chiral level. In this regime, the magnetization is positive and
increases as $1/B^{2}$ while the susceptibility increases as $1/B^{3}$ (see
below where we derive these results). We denote the critical magnetic field
where the WSM enter the quantum limit by $B_{1}$ and study its behavior in
the next section.

To see the importance of the chiral level, we show (the green curve in Fig.
3)\ the behavior of the susceptibility when the chiral level is artifically
removed from the calculation. Note that, in this case, the first
discontinuity near $1/B\approx 0.4$ T$^{-1}$ corresponds to the transition
of the Fermi level from $n=1$ to $n=2$ and not from $n=0$ to $n=1$ as in
real WSM$.$ With no chiral level, the oscillations are phase shifted with
respect to those of a real WSM. Their large $B$ behavior is also different.
Without the chiral level, the susceptibility is positive instead of negative
at large $B$ as shown in the inset of Fig. \ref{fig3}. Moreover, at large $%
B, $ the electrons condense at the bottom of the $n=1$ level so that the
susceptibility $\chi _{m}\sim B^{3/2}$. The large $B$ behavior of $\chi _{m}$
in the WSM can also be contrasted with that of the three-dimensional Schr%
\"{o}dinger fermions where $\chi _{m}$ $\sim B^{-4}$.

The magnetization goes to zero at small $B$ in the absence of a tilt as
expected on physical grounds. When $\chi t_{z}<0,$ however, the
magnetization tends to a constant positive value $m_{0}$ at small $B$ and
inversely if $\chi t_{z}>0$ where it tends to $-m_{0}.$ In all cases,
however, the magnetization due to the added carriers increases linearly with 
$B$ at small $B$ and the magnetic susceptibility $\chi =dm/dB>0.$ The
response is paramagnetic. For a WSM with two nodes of opposite chirality,
the minimal number of nodes required by the Nielsen-Ninomiya theorem\cite%
{Nielsen}, both nodes would need to have the same tilt in order for the
magnetization to vanish in the $B\rightarrow 0$ limit. This is not possible,
however, if inversion symmetry is to be preserved since opposite tilts are
then required. There would thus be a spontaneous magnetization in this case.
To preserve time-reversal symmetry, at least four nodes are required and the
summation of $\chi t_{z}$ over these nodes gives zero hence no spontaneous
magnetization. This spontaneous magnetization has been discussed before (see
Ref. \onlinecite{Mikitik2019}).

We can consider a Dirac node as two Weyl points of opposite chiralities but
with the same tilt located at the same wave vector $\mathbf{k}_{0}$ in the
Brillouin zone. From the previous paragraph, the spontaneous magnetization
is then zero for a Dirac node. A Dirac node has two chiral levels ($n=0$)
with opposite chiralities and the Landau levels $n\neq 0$ are twofold
degenerate in spin. Apart from this degeneracy, these $n\neq 0$ levels have
the same dispersion than the Landau levels in a Weyl node (assuming no
energy bias). The Weyl node, however, has only one chiral level. The
different behavior with respect to the spontaneous magnetization thus comes
from the chiral level i.e. from the first term on the right-hand side of Eq.
(\ref{energy}). For a Weyl node, the energy of the electron gas in the $n=0$
level is $U_{W}=\zeta \frac{e_{F}^{2}/2}{\beta +\chi t_{z}}$ while for a
Dirac node it is $U_{D}=\zeta \frac{e_{F}^{2}/2}{\beta +\chi t_{z}}\left( 
\frac{1}{\beta +\chi t_{z}}+\frac{1}{\beta -\chi t_{z}}\right) .$ We can
write%
\begin{equation}
U_{W}=\frac{1}{2}U_{D}-\frac{1}{2}\chi t_{z}\zeta \frac{e_{F}^{2}/2}{\beta
^{2}-t_{z}^{2}},  \label{miki}
\end{equation}%
so that the magnetization of a Weyl node is half that of a Dirac node but
with a correction that depends on the product $\chi t_{z}.$ (We recover in
this way Eq. (36) of Ref. \onlinecite{Mikitik2019}.)

To obtain the magnetization of the WSM and not just that of the added
carriers, one must also consider the contribution of the filled states in
the valence band (the vacuum). This contribution has been studied in a
number of papers (for a review, see Ref. \onlinecite{Mikitik2019}). It is
found that the occupied states in the valence band are responsible for a
giant diamagnetic anomaly in the magnetic susceptibility which diverges as
the Fermi level goes to zero when $B\rightarrow 0$ i.e. $\chi _{m}\sim -\ln
\left( \frac{E_{c}}{E_{F}}\right) ,$ where $E_{c}$ is a high-energy cutoff.
Moreover, it has been shown\cite{Carbotte1} that, at zero tilt, the vacuum
gives a negative contribution to the magnetization which is linear in $B$
and so a negative contribution to the magnetic susceptibility. It does not
contribute to the magnetic oscillations, however. At the opposite, in the
extreme quantum limit where the magnetization due to the added carriers goes
to zero, the vacuum diamagnetic response will dominate the response of the
Weyl semimetal, giving a magnetization that increases without limit as $B$
increases. This is the so-called magnetic torque anomaly\cite{Moll}. (See
also the last paragraph in Sec. III where we comment more on this point.)

\subsection{Behavior of $B_{1}$ and the quantum limit}

\begin{figure}
\centering\includegraphics[width = \linewidth]{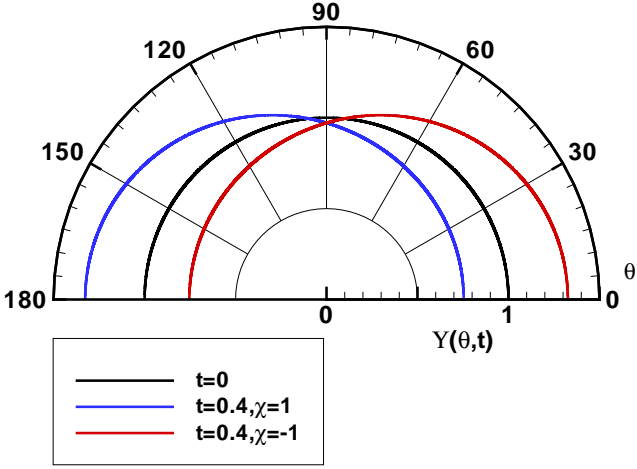} 
\caption{Angular dependence $\Upsilon
\left(t,\protect\theta \right) $ of $1/B_{1}$ for tilts $t=0$ and $t_{z}=0.4$ and both chiralities.}
\label{fig4}
\end{figure}

For a single node with chirality $\chi $ and tilt $\mathbf{t}$ filled with a
density of electrons $n_{e},$ the peaks in the oscillations of the physical
quantities occur each time the Fermi energy is at the bottom of an energy
level $n>0$ i.e. whenever $e_{F}=\min \left[ e_{n>0}\right] .$ From Eq. (\ref%
{densities}), the magnetic field at these particular values is given by:

\begin{equation}
\frac{1}{B_{n}}=\kappa \left( \mathbf{t}\right) F\left( n\right) ,
\label{bweyl}
\end{equation}%
where we have defined the function%
\begin{equation}
F\left( n\right) =\left[ \frac{\sqrt{n}}{\chi t_{z}+\beta }+\frac{2\beta }{%
\gamma }\sum_{n^{\prime }=1}^{n^{\prime }=n-1}\sqrt{n^{\prime }}\right]
^{2/3}  \label{f0}
\end{equation}%
and the parameter%
\begin{equation}
\kappa \left( \mathbf{t}\right) =\left( \frac{e}{\hslash }\right) \left( 
\frac{\left( 2\beta \gamma \right) ^{1/2}}{4\pi ^{2}n_{e}}\right)
^{2/3}=0.356\frac{\left( \beta \gamma \right) ^{1/3}}{\overline{n_{e}}^{2/3}}%
.
\end{equation}%
In particular, the transition of the Fermi level from the chiral level to $%
n=1$, i.e. the transition to the quantum limit, occurs at a magnetic field $%
B_{1}$ given by%
\begin{eqnarray}
\frac{1}{B_{1}} &=&\kappa \left( \mathbf{t}\right) \frac{1}{\left( \chi
t_{z}+\beta \right) ^{2/3}}  \label{bchiral} \\
&=&\frac{0.356}{\left( \overline{n_{e}}\right) ^{2/3}}\Upsilon \left(
t,\theta \right)  \notag
\end{eqnarray}%
where we have defined the function%
\begin{equation}
\Upsilon \left( t,\theta \right) =\frac{\left( \left( 1-t^{2}\right) \sqrt{%
1-t^{2}\sin ^{2}\theta }\right) ^{1/3}}{\left( \chi t\cos \theta +\sqrt{%
1-t^{2}\sin ^{2}\theta }\right) ^{2/3}}.
\end{equation}%
The quantum limit is reached at a smaller $B$ field when the density is
decreased. The angular dependence of the function $\Upsilon $ is shown in
Fig. \ref{fig4} for tilts $t=0$ and $t_{z}=0.4$ and for the two chiralities.
There is no angle dependence at zero tilt. The field $B_{1}$ can be measured
by torque magnetometry experiments\cite{Moll}.

\subsection{Periodicity of the oscillations in the $B\rightarrow 0$ limit}

\begin{figure}
\centering
\includegraphics[width = \linewidth]{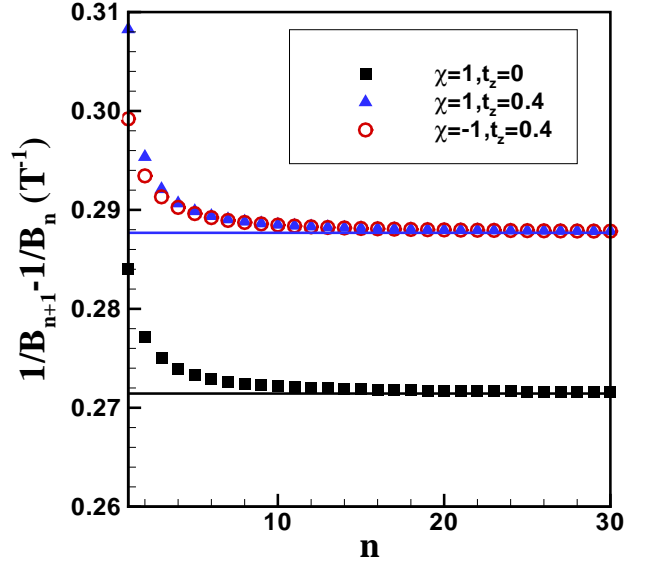} 
\caption{The function $P\left( \mathbf{t},n\right) $ as a function of $n$ for different values of the chirality and
tilt. The full lines gives the limit $\lim_{n\rightarrow \infty }P\left( 
\mathbf{t},\protect\chi ,n\right) .$} \label{fig5}
\end{figure}

If $B_{n}$ is the magnetic field where the Fermi level is just below level $%
n $ and $B_{n+1}$ where it is just below $n+1,$ then the separation between
two discontinuities in the slope of the oscillations is given by%
\begin{eqnarray}
P\left( \mathbf{t},n\right) &\equiv &\frac{1}{B_{n+1}}-\frac{1}{B_{n}} 
\notag \\
&=&\kappa \left( \mathbf{t}\right) \left[ F\left( n+1\right) -F\left(
n\right) \right] ,
\end{eqnarray}%
in units of Tesla$^{-1}.$ Figure \ref{fig5} shows that $P\left( \mathbf{t}%
,n\right) $ depends on $n.$ The oscillations contain multiple Fourier
components in $1/B,$ they are not periodic in $1/B$ in contrast with the
oscillations from two-dimensional Schr\"{o}dinger fermions. For large $n,$
however, Fig. \ref{fig5} indicates that $P\left( \mathbf{t},n\right) $ is
constant and we can write in this limit:%
\begin{equation}
\lim_{n\rightarrow \infty }F\left( n+1\right) -F\left( n\right) \rightarrow
2\left( \frac{2}{9}\right) ^{1/3}\left( \frac{2\beta }{\gamma }\right)
^{2/3}.
\end{equation}%
It is thus possible to define a period (in units of Tesla$^{-1}$) in this
small $B$ limit by%
\begin{eqnarray}
\lim_{n\rightarrow \infty }P\left( t,\theta ,n\right) &=&2\left( \frac{2}{3}%
\right) ^{2/3}\frac{e}{\hslash }  \label{periode} \\
&&\times \left( \frac{1}{4\pi ^{2}n_{e}}\right) ^{2/3}\Gamma \left( t,\theta
\right)  \notag \\
&=&0.430\,89\left( \frac{1}{\overline{n_{e}}}\right) ^{2/3}\Gamma \left(
t,\theta \right) ,  \notag
\end{eqnarray}

\begin{figure}
\centering
\includegraphics[width = \linewidth]{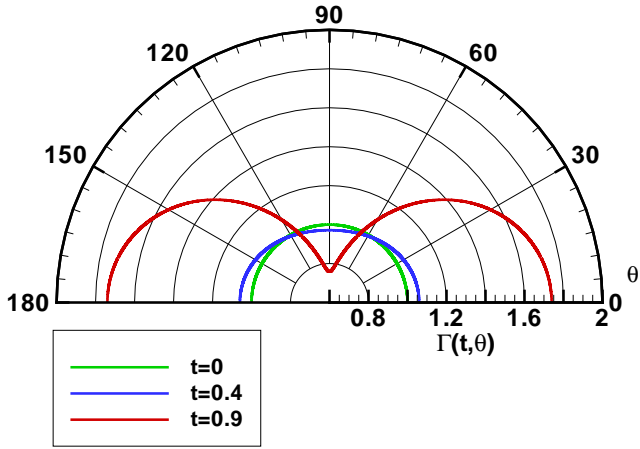} 
\caption{Angular dependence of the
function $\Gamma \left( \mathbf{t},\protect\theta \right) $ entering in the
fundamental period of the magnetic oscillations.} \label{fig6}
\end{figure}

\begin{equation}
\Gamma \left( t,\theta \right) =\frac{\sqrt{1-t^{2}\sin ^{2}\theta }}{\left(
1-t^{2}\right) ^{1/3}}
\end{equation}%
shows the anisotropy of the period. In the absence of tilt, this period $P$
is precisely that given by the dominant oscillatory term in the Poisson
formula for the magnetization\cite{Carbotte1} [if the chemical potential in
Eq. (38) of this reference is replaced by the $B=0$ result given by our Eq. (%
\ref{classical})]. With a tilt along $z,$ the Fermi surface becomes
ellipsoidal instead of spherical and $\lim_{n\rightarrow \infty }P\left(
t,\theta =0,n\right) =2\pi e/\hslash S$ is nothing but the usual De Haas-Van
Alphen period with $S$ the area in $\mathbf{k}$ space of the maximal orbit
for $\mathbf{B}$ along the $z$ direction. This period does not depend on the
chirality or on the sign of the tilt component $t_{z}$ or on the Fermi
velocity. It has the angular dependence shown in Fig. \ref{fig6}.

It is interesting to compare Eq. (\ref{bweyl}) with the corresponding
results for three-dimensional Schr\"{o}dinger's fermions which have the
dispersion 
\begin{equation}
E_{n}=\left( n+\frac{1}{2}\right) \hslash \omega _{c}+\frac{\hslash ^{2}k^{2}%
}{2m_{e}},
\end{equation}%
with $m_{e}$ the electron mass and $\omega _{c}=eB/m_{e}$ the cyclotron
frequency. A calculation following exactly the same steps as above gives in
the Schr\"{o}dinger case: 
\begin{equation}
\frac{1}{B_{n,S}}=2\frac{e}{\hslash }\left( \frac{1}{4\pi ^{2}n_{e}}\right)
^{2/3}\left( \sum_{n^{\prime }=1}^{n}\sqrt{n^{\prime }}\right) ^{2/3}
\end{equation}%
while for Weyl fermions with no tilt%
\begin{equation}
\frac{1}{B_{W,n}}=2\left( \frac{e}{\hslash }\right) \left( \frac{1}{4\pi
^{2}n_{e}}\right) ^{2/3}\left[ -\frac{1}{2}\sqrt{n}+\sum_{n^{\prime }=1}^{n}%
\sqrt{n^{\prime }}\right] ^{2/3}.
\end{equation}%
In the large $n$ limit, both expressions give the same period for the
oscillations, namely (setting $t=0$ for the Weyl node)%
\begin{equation}
\lim_{n\rightarrow \infty }P\left( n\right) =2\left( \frac{2}{3}\right)
^{2/3}\frac{e}{\hslash }\left( \frac{1}{4\pi ^{2}n_{e}}\right) ^{2/3}.
\end{equation}%
Moreover, in the large $n$ limit, we find the relation%
\begin{equation}
\frac{1}{B_{n,S}}\approx \frac{1}{2}\left[ \frac{1}{B_{n+1,W}}+\frac{1}{%
B_{n,W}}\right] ,
\end{equation}%
so that the Schr\"{o}dinger and Weyl oscillations are out of phase by half a
period as pointed out in Ref. \onlinecite{Carbotte1}.

\subsection{Magnetization and susceptibility in the quantum limit}

The quantum limit is reached when the magnetic field is such that the Fermi
level intersects only the chiral level i.e. $e_{F}\in \left[ 0,\min \left[
e_{1}\right] \right] $ for electron or $e_{F}\in \left[ \max \left[ e_{-1}%
\right] ,0\right] $ for holes. From Eq. (\ref{densities}), the Fermi level
is then given by%
\begin{equation}
e_{F}=4\pi ^{2}\ell ^{3}n_{e}\left( \chi t_{z}+\beta \right) .
\end{equation}%
It asymptotically approaches the neutrality point $e_{F}\rightarrow 0$ at
large $B$. With this expression in Eq. (\ref{energy}), the energy per
carrier in this limit is given by%
\begin{equation}
U=\frac{\left\vert n_{e}\right\vert h^{2}v_{F}}{2eB}\left( \chi t_{z}+\beta
\right)
\end{equation}%
and so the magnetization and susceptibility per carrier are given by%
\begin{equation}
m=\frac{\left\vert n_{e}\right\vert h^{2}v_{F}}{2\mu _{B}eB^{2}}\left( \chi
t_{z}+\beta \right)  \label{magne}
\end{equation}%
and%
\begin{equation}
\chi _{m}=-\frac{\left\vert n_{e}\right\vert h^{2}v_{F}}{\mu _{B}eB^{3}}%
\left( \chi t_{z}+\beta \right) .  \label{suscep}
\end{equation}

The magnetization of Weyl electrons is positive in this limit (since $\chi
t_{z}+\beta >0$) a behavior observed in the Weyl semimetal NbAs for example%
\cite{Moll}. It also goes to zero as $B\rightarrow \infty .$ This contrasts
with the behavior of Schr\"{o}dinger electrons in the quantum limit where
the magnetization per electron goes to the negative value $m=-1$ (in units
of $\mu _{B}$) at large $B.$

The susceptibility increases or decreases with respect to its value at zero
tilt depending on the sign of the product $\chi t_{z}.$ As we pointed out
above, one can show in the strong magnetic field limit that for a Weyl
semimetal the susceptibility $\chi _{m}\sim 1/B^{3/2}$ if the chiral level
is removed (see Fig. \ref{fig3}) while $\chi _{m}\sim -1/B^{4}$ for
three-dimensional Schr\"{o}dinger fermions and $\chi _{m}\sim 1/B^{3}$ for
Weyl fermions.

We remark that Eqs. (\ref{magne}-\ref{suscep}) are obtained by
differentiating the energy (or equivalently the Helmholtz free energy at $%
T=0 $ K) with respect to the magnetic field keeping the density constant.
Differentiation of the grand potential $\Omega $ at constant Fermi energy
(or chemical potential at $T=0$ K) gives, instead, in the extreme quantum
limit,%
\begin{equation}
m=\frac{1}{2}\frac{eE_{F}^{2}}{h^{2}v_{F}\mu _{B}\left( \chi t_{z}+\beta
\right) }
\end{equation}%
for the magnetization (in units of Bohr magneton per volume) and the
susceptibility is 
\begin{equation}
\chi _{m}=0.
\end{equation}%
Thus, when the Fermi level is kept constant and the WSM\ enters the extreme
quantum limit, the magnetic susceptibility goes to zero and the filled
states in the valence band dominate the magnetic response.

\section{ \ QUANTUM OSCILLATIONS FROM TWO\ WEYL NODES}

The Nielsen-Ninomiya theorem\cite{Nielsen} requires that the number of Weyl
points in the Brillouin zone be even so that Weyl nodes must occur in pairs
of opposite chirality. For simplicity, we analyse the quantum oscillations
due to a pair of nodes of opposite chirality and bias but with the same tilt
modulus $\left\vert \mathbf{t}\right\vert .$ We compute the total
magnetization and susceptibility for the two cases $t_{z,1}=\pm t_{z,2}$
(but the same value of $t_{\bot }$). We name these two cases WSM1 and WSM2.
Their parameters are defined in Tab. \ref{tab1}. In both cases, $\beta
_{1}=\beta _{2}=\beta ;\;\gamma _{1}=\gamma _{2}=\gamma $ where the
subscript here is the node index. For the numerical calculations, we take $%
n_{e}=2\times 10^{22}$ m$^{-3}$ for the\textit{\ total} electronic density
and $v_{F}=3\times 10^{5}$ m/s for the Fermi velocity. We define $t_{z}$ and 
$\Delta _{0}$ as positive. The energy scale is set by 
\begin{equation}
\frac{\hslash v_{F}}{\ell }=7.\,\allowbreak 70\sqrt{B}\,\text{meV.}
\end{equation}

\begin{table}[tbp] \centering%
\begin{tabular}{|l|l|}
\hline
\textbf{WSM1} & \textbf{WSM2} \\ \hline
$\chi _{1}=-\chi _{2}=1$ & $\chi _{1}=-\chi _{2}=1$ \\ \hline
$t_{z,1}=t_{z,2}=t_{z}$ & $t_{z,1}=-t_{z,2}=t_{z}$ \\ \hline
$\Delta _{0,1}=-\Delta _{0,2}=\Delta _{0}$ & $\Delta _{0,1}=-\Delta
_{0,2}=\Delta _{0}$ \\ \hline
\end{tabular}%
\caption{Parameters for the two-node Weyl semimetals 1 and 2.}\label{tab1}%
\end{table}%

We implicitly assume that the bias is not too large so that the two Weyl
nodes have separate Fermi surface. In real system, if the Fermi level lies
too far from the Dirac point, the two surfaces may merge into one surface
that encompasses both nodes.

If there were no scattering between the nodes, we would compute the common
Fermi level for some initial magnetic field $\mathbf{B}$ and find the
corresponding density of electrons in each node. Then as the magnetic field
is increased or decreased to study the quantum oscillations, the Fermi level
of the two nodes would differ but the electron density in each node will not
change. At large $B,$ the Fermi level $E_{F,i}$ in node $i$ will approach
the its neutrality point. Thus, for independent nodes, the total
susceptibility would simply be the sum of the susceptibility of each node.

For dependent nodes, scattering at finite temperature will modify the
density in each node so that they will always share the same Fermi level as
the magnetic field changes. In our calculations, we assume a finite doping
so that $E_{F}>\Delta _{0}$ initially. Upon increasing the magnetic field,
the common Fermi level can eventually cross the neutrality point in the node
with the positive bias thus creating holes in that node (i.e. a negative
electron density). The total density of electrons, however, must remain
constant. We study the case of dependent nodes which is the real physical
situation, for the rest of this section. We assume electron doping, i.e. $%
n_{e}>0.$

If the two nodes of WSM1 are located at the same wave vector, $\mathbf{k}%
_{0},$ and if there is no energy bias, then WSM1 can be considered as a node
of a Dirac semimetal while WSM2 (with the two nodes located at $\pm \mathbf{k%
}_{0}$) represent a Weyl semimal with space inversion symmetry. As we
mentionned above, at zero energy bias, the distinction between the two
metals as regards their magnetic behavior comes from the difference in the
chiral level.

\subsection{Density of states and ground state energy}

\begin{figure}
\centering
\includegraphics[width = \linewidth]{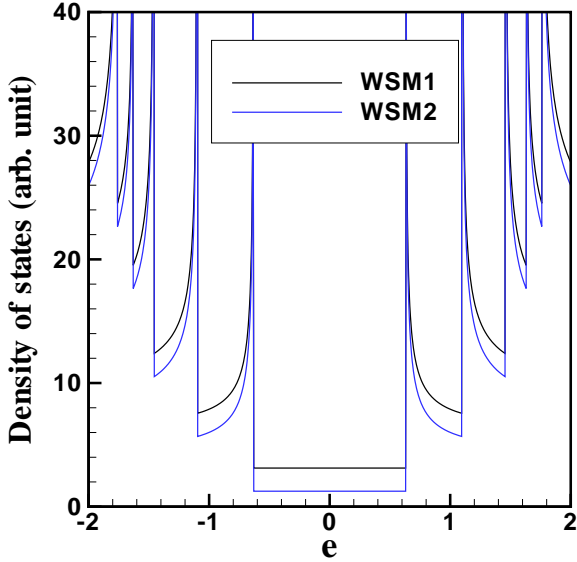} 
\caption{Density of states for the two
WSMs for bias $Q_{0}\ell =0.5$ and tilt $t_{z}=0.6.$} \label{fig7}
\end{figure}

Using Eqs. (\ref{dosnoeud}-\ref{dosnoeud2}), the density of states for the
two nodes in WSM1 and WSM2 can be written as 
\begin{eqnarray}
g_{1}\left( e\right) &=&g_{0,+}+g_{0,-}  \label{g1} \\
&&+g_{>}\left( e-Q_{0}\ell \right) +g_{>}\left( e+Q_{0}\ell \right) ,  \notag
\\
g_{2}\left( e\right) &=&2g_{0,+}+g_{>}\left( e-Q_{0}\ell \right)
+g_{>}\left( e+Q_{0}\ell \right) .  \label{g2}
\end{eqnarray}%
They differ by the constant%
\begin{equation}
g_{1}\left( e\right) -g_{2}\left( e\right) =g_{0,-}-g_{0,+}=\frac{2\alpha
t_{z}}{1-t^{2}}.
\end{equation}%
A finite tilt $t_{z}$ increases the density of states in WSM1 and decreases
it in WSM2. The difference between the two densities of states increases
rapidly with $t_{z}$. Figure \ref{fig7} shows the two densities of states
for $Q_{0}\ell =0.5$ and $t_{z}=0.6$ and a fixed magnetic field. Note that
the gap $\Delta e$ between the peaks at $n=-1$ and $n=1$ decreases as $%
\Delta e=2\sqrt{2}-2Q_{0}\ell $ with increasing bias. Equation (\ref{emin})
shows that the position in energy of the peaks in the density of states does
not depend on the chirality or sign of $t_{z}$ so that both densities of
states have the same structure in energy at any bias, apart from the shift
due to the chiral Landau level.

The Fermi level for either WSM is found by solving the equation%
\begin{equation}
n_{e}=\frac{\hslash v_{F}}{\ell }\int_{Q_{0}\ell }^{e_{F}}g_{1}\left(
e\right) de+\frac{\hslash v_{F}}{\ell }\int_{-Q_{0}\ell }^{e_{F}}g_{2}\left(
e\right) de,  \label{netotale}
\end{equation}%
and the total energy per electron is then given by%
\begin{eqnarray}
U &=&\frac{1}{n_{e}}\left( \frac{\hslash v_{F}}{\ell }\right)
^{2}\int_{Q_{0}\ell }^{e_{F}}g_{1}\left( e\right) ede  \label{epere} \\
&&+\frac{1}{n_{e}}\left( \frac{\hslash v_{F}}{\ell }\right)
^{2}\int_{-Q_{0}\ell }^{e_{F}}g_{2}\left( e\right) ede.  \notag
\end{eqnarray}

\subsection{Magnetic oscillations at zero tilt and finite bias}

\begin{figure*}[tbp]
\centering
\includegraphics[width = \linewidth]{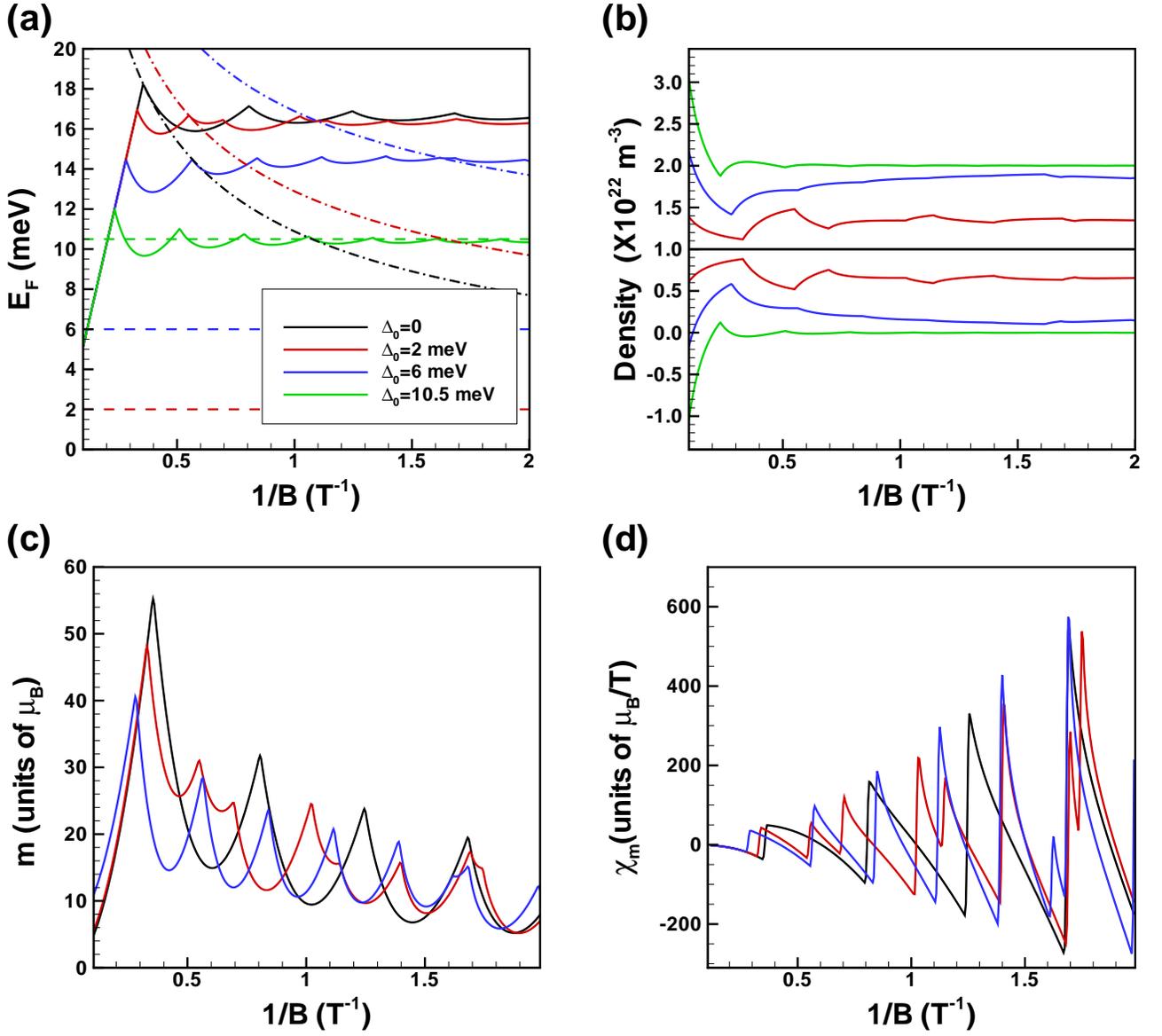}
\caption{Quantum oscillations as a function of the inverse magnetic field
for a WSM\ with zero tilt and for different values of the bias $\Delta _{0}:$
(a) Fermi level, (b) node densities, (c) magnetization and (d) magnetic
susceptibility. The dashed lines in (a)\ are set at the different values of $%
\Delta _{0}.$ The density is $n_{e}=2\times 10^{22}$ m$^{-3}.$}
\label{fig8}
\end{figure*}

Figure \ref{fig8} shows the oscillations in the Fermi level, node density,
magnetization and susceptibility, for different values of the bias, when $%
t=0 $ in which case there is no difference between the two WSMs and the
magnetization goes to zero at $B=0.$

As was the case for a single node, the discontinuities in the quantum
oscillations occur every times the magnetic field is such that the chemical
potential reaches the minimum of an energy band, i.e. whenever the condition%
\begin{equation}
e_{F}\left( B_{n}\right) =\min \left[ e_{n>0,\tau }\right] =Q_{0,\tau }\ell +%
\sqrt{2\beta _{\tau }\gamma _{\tau }n}  \label{efermi}
\end{equation}%
is satisfied for a given node $\tau $ and Landau level $n.$ The
corresponding magnetic field $B_{n}$ is found by solving%
\begin{eqnarray}
n_{e} &=&\frac{\hslash v_{F}}{\ell }\int_{Q_{0}\ell }^{\min \left[
e_{n>0,\tau }\right] }g_{1}\left( e\right) de  \label{densite} \\
&&+\frac{\hslash v_{F}}{\ell }\int_{-Q_{0}\ell }^{\min \left[ e_{n>0,\tau }%
\right] }g_{2}\left( e\right) de,  \notag
\end{eqnarray}%
where $e_{n>0,\tau }$ in the integration limit is an energy level of either
node since the Fermi level passes through many of them as the magnetic field
is varied.

In our calculation, we choose the density and bias such that the Fermi level
always satisfy $e_{F}>\max \left[ e_{n=-1,1}\right] $ so that we do not need
to consider the possibility that Landau levels $n\leq -1$ in node $1$ may be
occupied with holes. Holes may be present in the chiral level of node $1,$
however, when electrons are transferred to node 2. This happens when the
Fermi level $E_{F}$ drops below $\Delta _{0}$, a situation that occurs at $%
\Delta _{0}=10.5$ meV in Fig. \ref{fig8}(a). There is correspondingly a
negative density of electrons in node 1 as can be seen in the panel (b) of
this figure. The first peak in $1/B$ in Fig. \ref{fig8}(c) corresponds to $%
1/B_{1}$ for node $2$ for which $\Delta _{0}<0.$ This node has the largest
density of electrons and so reaches the quantum limit at a higher magnetic
field. The dashed lines in pannel (a) give the position of the Dirac point
in the left node while the dashed-dotted lines indicate the energy of the
Landau level $n=1,$ in the left node, below which the Fermi level enters the
quantum limit. For $\Delta _{0}=10.5$ meV, this node is always in the
quantum limit and the oscillations are due to the electrons in the second
node. The doubling of the peaks in panel (a) for $\Delta _{0}=2$ meV is a
clear indication that the system has not reached the quantum limit in either
node.

\begin{figure}
\centering
\includegraphics[width = \linewidth]{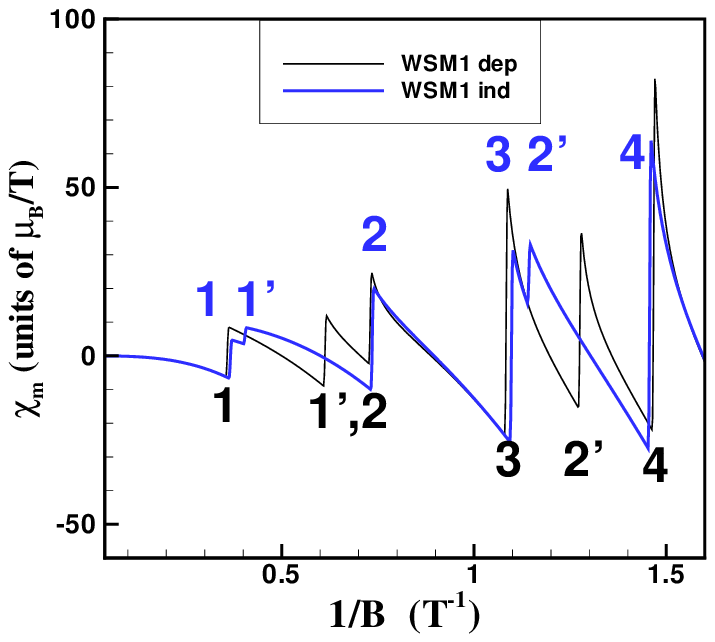} 
\caption{Susceptibility per carrier
calculated for WSM1 for dependent and independent nodes assuming the same
total density $n_{e}=2\times 10^{22}$ m$^{-3}$ , bias $\Delta _{0}=2$ meV
and tilt $t_{z}=0.5.$ The number above or below each peak indicates the
Landau level that is crossed by the Fermi level. Black (blue) numbers are
for dependent (independent) nodes and $n(n^{\prime })$ stand for node $2(1).$
Node 1(2) is shifted upward(downward) in energy (see Table 1).} \label{fig9}
\end{figure}

The pattern of oscillation changes if we include a tilt of the Weyl nodes in
addition to the bias and if we consider the nodes as independent instead of
as sharing a common Fermi level. We show an example of the difference
between dependent and independent nodes in Fig. \ref{fig9} for WSM1 with
bias $\Delta _{0}=2$ meV and tilt vector $t_{z}=0.5.$ In the independent
case, we calculate the initial position of the Fermi level at $B=0.5$ T,
assuming an equilibrium between the two nodes at that initial field. We
assume the same total density $n_{e}=2\times 10^{22}$ m$^{-3}$ in both cases.

The difference between dependent and independent nodes is more pronounced
when the WSM is compensated i.e. when there is initially an equal number of
electrons and holes. If the nodes are dependent, the Fermi level will not
move with a variation of the magnetic field since $n_{e}=0$ and so the
susceptibility will be zero (see Eqs. (\ref{sus1}-\ref{sus2}) below). For
WSM1, the Fermi level will be lying between $-\Delta _{0}$ and $+\Delta _{0}$
since $\left\vert t_{z}/\beta \right\vert <1$ while for WSM2, it will be
exactly at $E_{F}=0.$ For independent nodes, the susceptibility of each node
does not depend on the sign of the carrier and the susceptibility will be
twice that of a single node for the susceptibility per volume.

\subsection{Quantum limit at finite tilt and bias}

The quantum limit is reached when the Fermi level is in the chiral level of
both nodes. When this occurs, the behavior of the Fermi level with the
magnetic field is given by%
\begin{eqnarray}
E_{F,\text{WSM1}} &=&\frac{h^{2}v_{F}n_{e}\gamma }{2\beta e}\frac{1}{B}-%
\frac{\Delta _{0}t_{z}}{\beta }, \\
E_{F,\text{WSM2}} &=&\frac{h^{2}v_{F}n_{e}\left( \beta +t_{z}\right) }{2e}%
\frac{1}{B},
\end{eqnarray}%
and is linear in $1/B$ as shown in Fig. \ref{fig8}(a). When $B$ is very
large $E_{F,\text{WSM1}}\rightarrow -\frac{\Delta _{0}t_{z}}{\beta }$ and $%
E_{F,\text{WSM2}}\rightarrow 0$ i.e. the Fermi level asymptotically
approaches the neutrality point of each WSM$.$ At zero tilt, $%
E_{F}\rightarrow 0$ for both WSMs at large $B,$ in contrast to the case of
independent nodes (no scattering) where the Fermi level in each node
approaches the corresponding neutrality point $\pm \Delta _{0.}$

The total energy per carrier is given in this limit by

\begin{eqnarray}
U_{WSM1} &=&\frac{1}{2}\zeta \frac{e_{F}^{2}-\left( Q_{0}\ell \right) ^{2}}{%
\beta +t_{z}}+\frac{1}{2}\zeta \frac{e_{F}^{2}-\left( Q_{0}\ell \right) ^{2}%
}{\beta -t_{z}}  \label{ener1} \\
&=&\frac{ev_{F}}{4\pi ^{2}n_{e}}\frac{1}{\beta }  \notag \\
&&\times \left( \frac{4\pi ^{4}\hslash ^{2}n_{e}^{2}\gamma }{e^{2}B}-\frac{%
4\pi ^{2}\hslash n_{e}Q_{0}t_{z}}{e}-BQ_{0}^{2}\right)  \notag
\end{eqnarray}%
and 
\begin{eqnarray}
U_{WSM2} &=&\frac{1}{2}\zeta \frac{e_{F}^{2}-\left( Q_{0}\ell \right) ^{2}}{%
\beta +t_{z}}+\frac{1}{2}\zeta \frac{e_{F}^{2}-\left( Q_{0,\tau }\ell
\right) ^{2}}{\beta +t_{z}}  \label{ener2} \\
&=&\frac{1}{2}\frac{ev_{F}}{4\pi ^{2}n_{e}}\left( \frac{2}{\beta +t_{z}}%
\right)  \notag \\
&&\times \left( \frac{4\pi ^{4}\hslash ^{2}n_{e}^{2}\left( \beta
+t_{z}\right) ^{2}}{e^{2}B}-BQ_{0}^{2}\right) .  \notag
\end{eqnarray}%
Equations (\ref{mag},\ref{sus}) give for the magnetization per carrier
\qquad 
\begin{eqnarray}
m_{WSM1} &=&\frac{h^{2}n_{e}v_{F}}{4\mu _{B}eB^{2}}\frac{\gamma }{\beta }+%
\frac{e}{\mu _{B}h^{2}n_{e}v_{F}}\frac{\Delta _{0}^{2}}{\beta },
\label{aimant1} \\
m_{WSM2} &=&\frac{h^{2}n_{e}v_{F}}{4\mu _{B}eB^{2}}\left( \beta +t_{z}\right)
\label{aimant2} \\
&&+\frac{e}{\mu _{B}h^{2}n_{e}v_{F}}\frac{\Delta _{0}^{2}}{\beta +t_{z}}, 
\notag
\end{eqnarray}%
and for the susceptibility per carrier

\begin{eqnarray}
\chi _{m,WSM1} &=&-\frac{h^{2}n_{e}v_{F}}{2\mu _{B}eB^{3}}\frac{\gamma }{%
\beta },  \label{chi1} \\
\chi _{m,WSM2} &=&-\frac{h^{2}n_{e}v_{F}}{2\mu _{B}eB^{3}}\left( \beta
+t_{z}\right) .  \label{chi2}
\end{eqnarray}%
When the Fermi level is in the chiral level of node 2, the susceptibility of
the two WSMs are independent of the bias. Moreover, the two WSMs then differ
only in their dependence on the tilt direction which is given by%
\begin{eqnarray}
\chi _{m,WSM1} &\sim &\left( 1-t^{2}\right) /\sqrt{1-t^{2}\sin ^{2}\theta },
\label{sus1} \\
\chi _{m,WSM2} &\sim &\sqrt{1-t^{2}\sin ^{2}\theta }+t\cos \theta .
\label{sus2}
\end{eqnarray}%
The $1/B^{2}$ behavior of the magnetization is clearly visible in Fig. \ref%
{fig8}(c)$.$ When only the chiral level is occupied, our calculation shows
that the susceptibility is negative at large $B$ and there is a constant
contribution to the magnetization at finite bias. This constant is very
small. At zero tilt, for example, it is given by%
\begin{equation}
m=\frac{e\Delta _{0}^{2}}{\mu _{B}h^{2}n_{e}v_{F}}=8.\,\allowbreak
528\,5\times 10^{-3}\frac{\overline{\Delta }_{0}^{2}}{\overline{n_{e}}}
\end{equation}%
in Bohr magneton per electron.

\subsection{Behavior of $B_{1}$ and periodicity of the oscillations at
finite tilt and bias}

The first peak at small $1/B$ occurs when the Fermi level $e_{F}\left(
B\right) =\min \left[ e_{1,2}\right] $ i.e. when the system enters the
quantum limit. It is then in the chiral level of both nodes so that only the
contribution to the density of states of these levels need to be considered.
The magnetic lengths $\ell _{1}$ and $\ell _{2}$ (corresponding to $1/B_{1}$%
) for WSM1 and WSM2 are given by solving the equations 
\begin{eqnarray}
\ell _{1}^{3}+\frac{Q_{0}}{\xi }\ell _{1}-\frac{1}{2\pi ^{2}n_{e}}\sqrt{%
\frac{2\beta ^{3}}{\gamma }} &=&0,  \label{l1} \\
\ell _{2}^{3}+\frac{Q_{0}}{\xi }\ell _{2}-\frac{\sqrt{2\beta \gamma }}{\xi }
&=&0,  \label{l2}
\end{eqnarray}%
where $Q_{0}\geq 0$ and we have defined the constant%
\begin{equation}
\xi =2\pi ^{2}n_{e}\left( \beta +t_{z}\right) .
\end{equation}%
If there is no tilt, the magnetic length at this peak is instead given by
the solution of the equation%
\begin{equation}
\ell _{0}^{3}+\frac{Q_{0}}{2\pi ^{2}n_{e}}\ell _{0}-\frac{1}{\sqrt{2}\pi
^{2}n_{e}}=0.  \label{qo}
\end{equation}%
In particular, at zero bias the position in $1/B$ of the first peak is%
\begin{equation}
\frac{1}{B_{1}}=0.564\,62\frac{1}{\overline{n}_{e}^{2/3}}\text{ T}^{-1},
\end{equation}%
which is simply Eq. (\ref{bchiral}) with a electronic density $n_{e}/2.$

\subsection{Magnetic oscillations and quantum limit at zero bias}

\begin{figure}
\centering
\includegraphics[width = \linewidth]{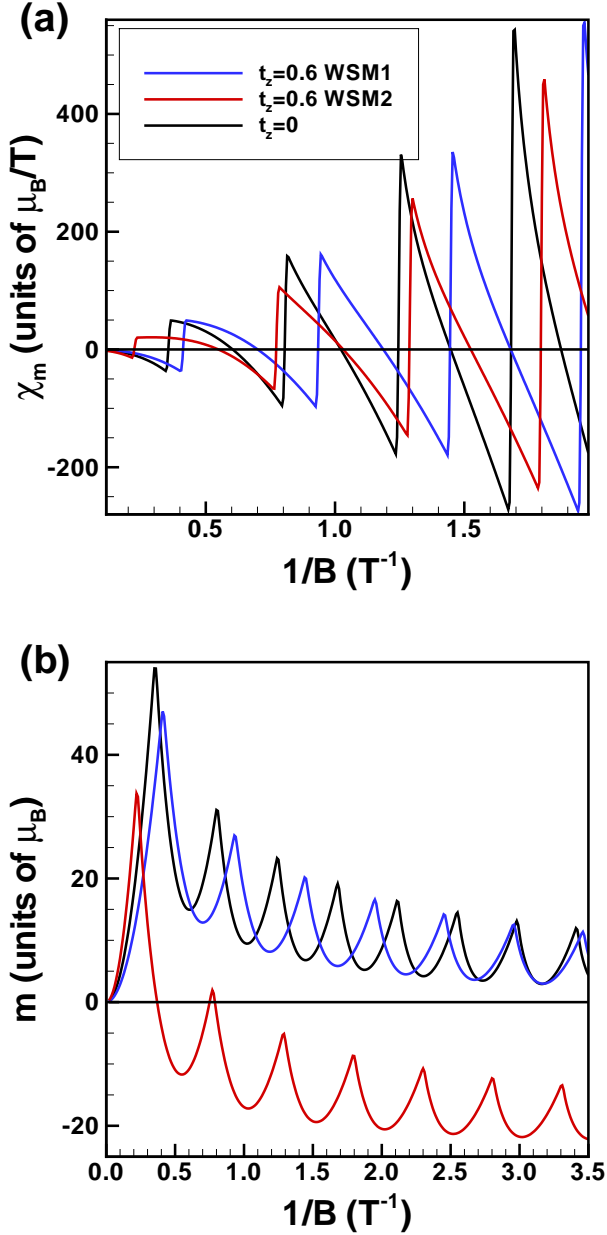} 
\caption{Effect of a finite tilt on (a)
the magnetic susceptibility and (b) magnetization of both WSMs at zero bias$.
$} \label{fig10}
\end{figure}

Figure \ref{fig10} shows the effect of a finite $t_{z}$ on the magnetic
susceptibility and magnetization of both WSMs for zero bias. The spacing
between the oscillations increases with $t_{z}$ for both WSMs while it
decreases with a finite $t_{\bot }$ (not shown in the figure). The
susceptibility decreases with $t_{z},$ more so for WSM2 than for WSM1. As
discussed in Sec. III, the magnetization does not go to zero at small $B$
for WSM2 since the two nodes have $\chi t_{z}=1$.

At zero bias, Eq. (\ref{periode}) can be generalized for WSM2 (opposite
tilts) to

\begin{equation}
\lim_{n_{0}\rightarrow \infty }\lambda \left( \mathbf{t},n_{0}\right)
=2\left( \frac{2}{3}\right) ^{4/3}\frac{e}{\hslash }\left( \frac{1}{4\pi
^{2}n_{e}}\right) ^{2/3}\left( \frac{\beta }{\gamma ^{1/3}}\right) ,
\end{equation}%
taking into account that, in this case, the node density is $n_{e}/2.$

Figure \ref{fig11} shows $1/B_{1}$ for both WSMs as a function of the polar
angle $\theta $ for different values of the bias $\Delta _{0}$ and tilt
modulus $t.$ If there is no tilt, there is no distinction between the two
WSMs at any bias. For a finite tilt, $1/B_{1}$ (WSM1)$>1/B_{1}$ (WSM2) if $%
t_{z}>0$ (i.e. $\theta <\pi /2$) and vice versa$.$ Both peaks are shifted to
lower values of $1/B$ by a finite bias. A finite tilt thus introduces a
dephasing that is different for the two Weyl semimetals and which is also
anisotropic.

\begin{figure}
\centering
\includegraphics[width = \linewidth]{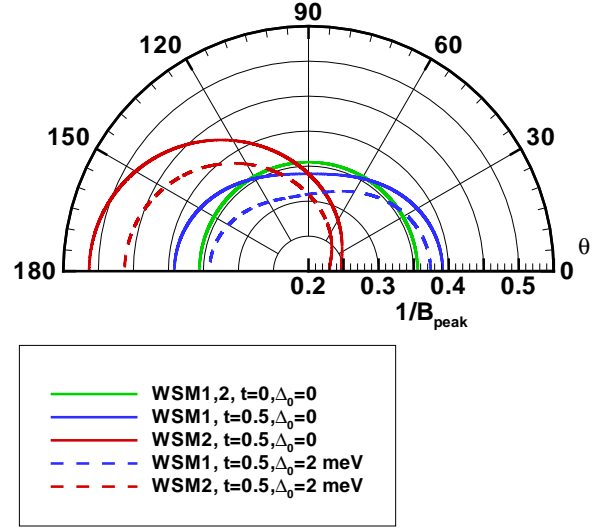} 
\caption{Position of the
first peak in $1/B$ of the quantum oscillations for WSM1 and WSM2.} \label%
{fig11}
\end{figure}

At zero bias, we can simplify Eq. (\ref{densite}) by using Eqs. (\ref{g1}-%
\ref{g2}) with $Q_{0}\ell =0.$ We get for the density 
\begin{equation}
n_{e}=\alpha \int_{0}^{\sqrt{2\beta \gamma n}}f\left( e\right) de,
\end{equation}%
where we have defined $f\left( e\right) =g\left( e\right) /\alpha $ and use
the fact that $\beta _{\tau },\gamma _{\tau }$ have the same value for both
nodes. For WSM1 and WSM2, this gives for the magnetic field at the peak $n$%
\begin{eqnarray}
\ell _{1}^{3}\left( n\right) &=&\frac{1}{4\pi ^{2}n_{e}}\left(
f_{0,+}+f_{0,-}\right) \sqrt{2\beta \gamma n} \\
&&+\frac{1}{2\pi ^{2}n_{e}}\int_{\sqrt{2\beta \gamma }}^{\sqrt{2\beta \gamma
n}}f_{>}\left( e\right) de  \notag
\end{eqnarray}%
and%
\begin{equation}
\ell _{2}^{3}\left( n\right) =\frac{1}{2\pi ^{2}n_{e}}\left[ f_{0,+}\sqrt{%
2\beta \gamma n}+\int_{\sqrt{2\beta \gamma }}^{\sqrt{2\beta \gamma n}%
}f_{>}\left( e\right) de\right]
\end{equation}%
with the definition%
\begin{equation}
f_{0,\pm }=\frac{1}{\beta \pm t_{z}}.
\end{equation}%
We thus find for the dephasing between the oscillations of the two WSMs the
relation,%
\begin{eqnarray}
\frac{1}{B_{1}^{3/2}} &=&\frac{1}{B_{2}^{3/2}}+\frac{1}{4\pi ^{2}n_{e}}%
\left( \frac{2e}{\hslash }\right) ^{3/2} \\
&&\times \left[ \frac{t\sqrt{n}\cos \theta }{\sqrt{1-t^{2}}}\left(
1-t^{2}\sin ^{2}\theta \right) ^{1/4}\right]  \notag \\
&=&\frac{1}{B_{2}^{3/2}}+0.424\,\frac{1}{\overline{n_{e}}}\frac{t\sqrt{n}%
\cos \theta }{\sqrt{1-t^{2}}} \\
&&\times \left( 1-t^{2}\sin ^{2}\theta \right) ^{1/4}  \notag
\end{eqnarray}%
Hence, the dephasing increases with the Landau level index $n$ and with the
tilt $t.$

\section{CONCLUSION}

In this paper, we have studied the contribution of the added carriers
(electron or hole) to the orbital magnetization and magnetic susceptibility
of a simple two-node model of a Weyl semimetal. We have studied how the
behavior of the quantum (de Haas-van Alphen) oscillations of the
magnetization and magnetic susceptibility is modified by a tilt of the Weyl
nodes and, considering a pair of nodes with opposite chirality, how these
oscillations change when both nodes have the same or opposite value of the
component of the tilt vector along the magnetic field direction. We have
also considered the effect of an energy bias between the two nodes.
Throughout our study, we emphasized the importance of the chiral level in
distinguishing the magnetic oscillations of Weyl semimetals from those of
Schr\"{o}dinger fermions or between Weyl and Dirac fermions. We discussed
the anisotropic behavior induced by the tilt vector in the fundamental
period of oscillation and in the magnetic field $B_{1}$ needed to reach the
quantum limit. Finally, we showed the difference in the quantum oscillations
between two nodes with and without internode scattering.

As we were concerned with the role of the added carriers in the magnetic
properties, we did not include the contribution of the filled states in the
valence band (the vacuum). Although they do not affect the magnetic
oscillations, they contribute to the magnetization and are required to
understand the magnetic torque anomaly at large magnetic field as well as
the giant diamagnetic anomaly at small magnetic field when the Fermi level
is close to the neutrality point.

Our simple model cannot, of course, reproduce the experimental results for
real Weyl semimetals. In real WSM, there may be different types of Fermi
surface pockets, both trivial and non-trivial (topological) which contribute
to the magnetic oscillations\cite{Arnold}. Moreover, the Fermi velocity and
so the Fermi surface may be anisotropic so that the period will depend in
general on the orientation of the magnetic field with respect to the
crystallographic axis. The energy bias and tilt of the different nodes at
the Fermi energy may differ. Finally, the Fermi arcs may contribute to the
magnetization.

The magnetic susceptibility of a single Weyl (or Dirac) node in the
continuum (linear) approximation that we use can be compared with that
obtained from a lattice model where the bandwiths are finite. Such a
comparison is made in Ref. \onlinecite{Koshino2016} where it is confirmed
that the continuum approximation is quite good if, as expected, the Fermi
level is not too far from the Dirac point.

\begin{acknowledgments}
R. C. was supported by a grant from the Natural Sciences and Engineering
Research Council of Canada (NSERC). S. V. was supported by a scholarship
from NSERC and FRQNT. Computer time was provided by Calcul Qu\'{e}bec and
Compute Canada.
\end{acknowledgments}

\end{document}